\documentclass[reprint,aps,prappl,showpacs,twocolumn]{revtex4}%
\usepackage{amsmath}
\usepackage{amsfonts}
\usepackage{amssymb}
\usepackage{graphicx}
\usepackage[colorlinks,linkcolor=blue,anchorcolor=blue,citecolor=blue,urlcolor=black]%
{hyperref}
\usepackage{mathrsfs}
\usepackage{dcolumn}
\usepackage{bm}
\usepackage{epsfig}%
\providecommand{\U}[1]{\protect\rule{.1in}{.1in}}

\def\be{\begin{equation}}
\def\ee{\end{equation}}
\begin{document}
\title{\bf Momentum Analysis for Metasurfaces}
\author{Wenwei Liu$^{1}$}
\author{Zhancheng Li$^{1}$}
\author{Hua Cheng$^{1}$}
\author{Shuqi Chen$^{1}$}
\email{schen@nankai.edu.cn}
\altaffiliation{URL: http://phy.nankai.edu.cn/grzy/schen/English/index.htm}
\author{Jianguo Tian$^{1}$}
\affiliation{$^{1}$The Key Laboratory of Weak Light Nonlinear Photonics, Ministry of Education, School of Physics and TEDA Institute of Applied Physics, Nankai University, Tianjin 300071, China.}
\date{\today}

\begin{abstract}
Utilizing discrete phase distribution to fit continuous phase distribution has been a primary routine for designing metasurfaces. In the existing method, the validation of the discrete designs is guaranteed only by using the sub-wavelength condition of unit cells, which is insufficient, especially for arbitrary phase distribution. Herein, we proposed an analytical method to design metasurfaces via estimating the width of the source in a unit cell. Also, by calculating field patterns in both real- and momentum-space, we provided four guidelines to direct future applications of metasurfaces, such as an arbitrary multi-foci lens with the same strength of each focus, a convex–concave double lens, and a lens with a large numerical aperture that can precisely prevent undesired diffraction orders. Besides metalens, this methodology can provide a wide platform for designing tailored and multifunctional metasurfaces in future, especially large-area ones in practical applications.
\end{abstract}

\maketitle

\section{INTRODUCTION}
Artificially engineered metasurfaces, comprising a dense arrangement of sub-wavelength resonators, are efficient and facilely fabricated substitution of bulk metamaterials \cite{1,2,3,4,5}. By introducing an abrupt phase shift to the incident wavefront, metasurfaces modify the scattered wavefront in deep sub-wavelength scale, such as anomalous refraction \cite{6,7}, metalenses \cite{8,9,10}, holographic plates \cite{11,12,13}, coding metasurfaces \cite{14}, wave plates \cite{15}, and asymmetric transmission \cite{16,17,18}. Other novel applications including spin Hall Effect \cite{19}, topological transitions \cite{20,21}, and nonlinear responses \cite{22,23} are also proposed with specific design of metasurfaces. It has been demonstrated that the generalized refraction is equivalent to blazed diffraction gratings \cite{6,24}. However, the existing quantitative diffracting theories mainly handle either a linearly distributed phase profile or holography, which is highly insufficient compared with the modulation depth of the metasurfaces. In addition, these theories always treat the discontinuous phase distribution as quasi-continuous, which is valid for most of the existing metasurfaces. However, when the phase of each unit cell varies abruptly, or the size of the unit cell is close to or even larger than the wavelength of the signals in nonlinear metasurfaces \cite{22}, the hypothesis of quasi-continuous is no longer validated.

Recently, metasurface holography enabling arbitrary wavefront reconstruction and optical communication has attracted considerable research interests in the scientific community \cite{25,26}. Unlike the traditional holograms, which is generated by interference of a reference beam with the scattered beam from a real object, the metasurface hologram is generated by numerically computing the phase information at the hologram interface using the computer-generated holography (CGH) method \cite{27}. With delicately controlled geometry of the antennas, the desired phase profile can be achieved to accomplish three-dimensional holography \cite{28}, surface plasmon holography \cite{29}, multiwavelength achromatic holography \cite{30}, and nonlinear holography \cite{31}. Using reflective−type plasmonic metasurfaces \cite{32} and dielectric Huygens’ metasurface \cite{33}, high-efficiency holograms can also be realized. It is known that holography is based on Fourier analysis to achieve information storage or image reconstruction. Fourier analysis is indeed a more fundamental theory, that can describe the scattered diffraction field for most of the metasurfaces besides holographic metasurfaces. However, the corresponding analyzing method has not been applied to an arbitrary metasurface, such as metalens.

Moreover, the reflection's law, regular or generalized Snell's law, and the grating equation are all derived from a single principle: conservation of momentum along the surface of the device (Demonstrated in Appendix A). Momentum, which is wave vector in electrodynamics, as well as energy are always dominant quantities. Herein, we provide a general guide to evaluate the diffracting field emitted from an arbitrarily arranged metasurface, with tailored functionalities but without considering the specific nanostructures or material details. First, we proposed a localized hypothesis of the unit cells composing metasurfaces. The sizes of source in the unit cells are in consideration compared with other works, in which the unit cells are just characterized by a point source with an effective dipole moment \cite{34} or multipolar components \cite{35}. Second, we employed the Fourier analysis to derive a generic wavefront in \emph{k}-space. We provided four guidelines for designing metasurfaces based on this momentum analyzing method, the most important one of which is the stable condition that fulfills the conditions to mimic continuous phase distribution using a discontinuous phase profile. Taking metalens as an example, we realized an arbitrary periodic-foci lens, a convex-concave double lens, and a large numerical aperture (NA) lens preventing undesired diffracting orders via the derived formulas. Our results provide a powerful tool to design the advanced functional and tunable metadevices, especially large-area ones in practical applications.

\section{MODELING APPROACH}

\begin{figure}[tb]
\centerline{\includegraphics[width=\columnwidth]{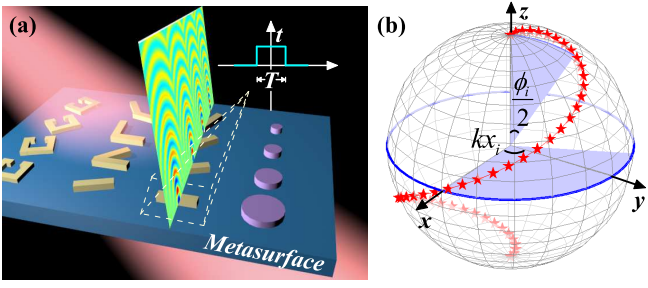}}
\caption{(color online) Localized rectangular model of diffraction theory for metasurfaces. (a) Schematic of an arbitrarily designed metasurface with radiation from every single unit cell. Inset: Responding function of the \emph{i}th unit cell with $t = \left| {{t_i}} \right|{e^{ - i{\phi _i}}}$ and $T = {T_i^{MS}}$.  (b) Sphere of the meta-unit with polar angle ${\phi _i}/2$ and azimuthal angle $k{x_i}$. The red starred line depicts a class of equivalence with ${\phi _i} + k{x_i} = \pi $.}
\label{fig1}%
\end{figure}

The strength and phase of output light are two main parameters that facilitate the design of tailorable metasurfaces. When any two adjacent unit cells are weakly−coupling (UCWC), these two parameters can characterize an independent unit cell regardless of the radiation type of the nanostructure \cite{6,7,8,9,10,11,12,13,14,15,16,17,18,28,29,30,31,32,33,36}. Based on this designing strategy, the responding function of a unit cell can be mathematically defined as a rectangular function $t(x) = \left| t \right|{\rm{rect}}(x/{T^{MS}}){e^{ - i\phi }}$, where $\left| t \right|$ is the responding strength (reflection or transmission) and $\phi$ is the phase delay of the unit cell. The factor ${\rm{rect}}(x/{T^{MS}})$ is the estimation of the locality for the unit cell, where ${T^{MS}}$ describes an equivalent size that the nanostructure can govern, as illustrated in Fig. \ref{fig1}(a). When the parameter of width ${T^{MS}}$ is small enough comparing with the resonant wavelength, the radiating nanostructure can be treated as a point source. Generally, the actual value of ${T^{MS}}$ should be decided via the geometry and materials of the nanostructures in a unit cell, and can be simply estimated by the size of the nanostructure. In Appendix B, we compared our theory with the experimental results in Ref. \cite{6} and simulated results in Ref. \cite{37}, where the ${T^{MS}}$ is set as the size of the nanostructures.
 The far field diffraction pattern is the Fourier transform of the responding function based on the principle of superposition \cite{38}: $\mathcal{F}(k) = \frac{{\left| t \right|}}{\pi }\frac{{\sin (k{T^{MS}}/2)}}{k}{e^{ - i\phi }}$. When the center of the unit cell is located at an arbitrary location ${x_{i}}$, $t(x)$ can be written as:
\begin{align}
{t_i}(x) = \left| {{t_i}} \right|{\rm{rect}}(\frac{{x - {x_i}}}{{{T_i^{MS}}}}){e^{ - i{\phi _i}}}.
\label{eq:e1}
\end{align}
According to the translation formula $\mathcal{F}[f(x - {x_0})] = {e^{ - ik{x_0}}}\mathcal{F}[f(x)]$, the Fourier transform of Eq. (\ref{eq:e1}) is:
\begin{align}
\mathcal{F}_i(k) = \frac{{\left| {{t_i}} \right|}}{\pi }\frac{{\sin (k{T_i^{MS}}/2)}}{k}{e^{ - i({\phi _i} + k{x_i})}}.
\label{eq:e2}
\end{align}
Equation (\ref{eq:e2}) can be illustrated in a meta-unit [Fig. \ref{fig1}(b)]. The radius of the sphere is $\frac{{\left| {{t_i}} \right|}}{\pi }\frac{{\sin (k{T_i^{MS}}/2)}}{k}$, characterizing the promoted responding strength of the unit cell. The polar angle ${\phi _i}/2$ and azimuthal angle $k{x_i}$ are decided by the phase delay and location of the unit cell, respectively. The meta-unit describes exactly how each unit cell functions in the total scattered wavefront. When ${\phi _i} + k{x_i}$ is a constant, as depicted by the red starred line in Fig. \ref{fig1}(b), $\mathcal{F}_i$ remains the same, which implies that all the units on the line are equivalent to each other despite their different phase delay and locations. This characteristic can be utilized to achieve phase retardation only by varying the location of a unit cell. It should be noted that ${\phi _i} + k{x_i}=\rm{C}$ also represents the wavefront (equiphase surface) of $\mathcal{F}_i$ in $k$-space; the envelope of which is a plane wave. The envelope is ${\phi _i} + {\bf{k}} \cdot {\bf{r}}_i = \rm{C}$ for two-dimensional metasurfaces, which remains a plane wave in \emph{k}-space (see Appendix C). Specifically, when the constant $\rm{C}$ of each unit remains the same, the equation for the neighboring unit cell is ${\phi_i}^\prime  + k{x_i}^\prime  = \rm{C}$. Thus, we obtain $k =  - ({\phi _i}^\prime  - {\phi _i})/({x_i}^\prime  - {x_i})$.
Considering the $2n\pi $ uncertainty of phase, the wavefront of $\mathcal{F}_i$ (defined as the localized wave vector) can be expressed as:
\begin{align}
k =  - \frac{{\Delta {\phi _i}}}{{\Delta {x_i}}} + \frac{{2n\pi }}{{\Delta {x_i}}}.
\label{eq:e3}
\end{align}
When $\phi_i$ varies linearly with $x$, Eq. (\ref{eq:e3}) is exactly the generalized Snell's law with a diffracting order of $n$. Furthermore, for a metasurface possessing finite or infinite unit cells, the total responding functions should be the sum of each cell (see Appendix C):
\begin{align}
T({\bf{r}}) = \sum\limits_i {\left| {{t_i}} \right|\prod\limits_{v = x,y}^{} {{\rm{rect}}(\frac{{v - {v_i}}}{{{T_{vi}^{MS}}}})} {e^{ - i{\phi _i}}}},
\label{eq:e4}
\end{align}
\begin{align}
\mathcal{F}({\bf{k}}) = \sum\limits_i {\frac{{\left| {{t_i}} \right|}}{\pi }{e^{ - i({\phi _i} + {\bf{k}} \cdot {{\bf{r}}_i})}}\prod\limits_{v = x,y}^{} {\frac{{\sin ({k_v}{T_{vi}^{MS}}/2)}}{{{k_v}}}} }.
\label{eq:e5}
\end{align}
Equations (\ref{eq:e4}) and (\ref{eq:e5}) express the far field diffraction patterns for all UCWC metasurfaces. Specifically, the diffracting field of a metasurface $M_1$ with $N$ unit cells is $\mathcal{F}(k) = \sum\limits_{i = 1}^N {\frac{{\left| {{t_i}} \right|}}{\pi }\frac{{\sin (k{T_i^{MS}}/2)}}{k}{e^{ - i({\phi _i} + k{x_i})}}}$. For simplicity, $\left| {{t_i}} \right| \equiv 1$ and ${T_i^{MS}} \equiv {T^{MS}} \ll \lambda$ are assumed. Then,
\begin{align}
\mathcal{F}(k) \propto {\mathcal{F}_1}(k) \equiv \sum\limits_{i = 1}^N {{e^{ - i({\phi _i} + k{x_i})}}}.
\label{eq:6}
\end{align}
Consider another metasurface $M_2$ with a diffracting field of
\begin{align}
\mathcal{F}_2(k) \equiv \sum\limits_{i = 1}^{N/2} {{e^{ - i({\phi _{2i}} + k{x_{2i}})}}}.
\label{eq:7}
\end{align}
$M_2$ is obviously composed of all the even unit cells in $M_1$. If the diffracting field of $M_2$ converges, the field should be analogous to the diffracting field of $M_1$, which implies that
\begin{align}
\mathcal{F}_2(k) \propto {\mathcal{F}_1}(k).
\label{eq:8}
\end{align}
On the other hand, $\mathcal{F}_1(k)$ and $\mathcal{F}_2(k)$ are related mathematically by
\begin{align}
\sum\limits_{i = 1}^N {{e^{ - i({\phi _i} + k{x_i})}}} = \sum\limits_{i = 1}^{N/2} {(1 + {e^{i\Delta {\phi _{2i}}}}{e^{ik\Delta {x_{2i}}}}){e^{ - i({\phi _{2i}} + k{x_{2i}})}}},
\label{eq:9}
\end{align}
where $\Delta {\phi _{2i}} = \phi {}_{2i} - \phi {}_{2i - 1}$, $\Delta {x_{2i}} = {x_{2i}} - {x_{2i - 1}}$.

Comparing Eqs. (\ref{eq:6})--(\ref{eq:9}), the Eq. (\ref{eq:8}) can be satisfied only when the coefficients of $1 + {e^{i\Delta {\phi _{2i}}}}{e^{ik\Delta {x_{2i}}}}$ in Eq. (\ref{eq:9}) equal to a constant for every $i$ and $k$. Thus, we attain the conditions for discrete phase distribution to mimic a continuous phase distribution:
\begin{align}
k\Delta {x_{2i}} \ll 2\pi,
\label{eq:10}
\end{align}
\begin{align}
\Delta {\phi _{2i}} \simeq \rm{C}.
\label{eq:11}
\end{align}
Generally, the subscript $2i$ can be replaced by $i$. The Eq. (\ref{eq:10}) is exactly the sub-wavelength condition, which can be simply written as $\Delta x \ll \lambda$. When the phase of each unit cell varies rapidly, Eq. \ref{eq:11} will be nullified, and the sub-wavelength condition will not guarantee validation of mimicking a continuous phase distribution. Specifically, $\Delta {\phi _i} \simeq 0$ means phase of the unit cell varies slowly. Equation (\ref{eq:11}) is called as stable condition in this study.

 For an arbitrary phase distribution, the following four issues are important:

1.	One-to-one correspondence. Each unit cell produces a plane wave in $k$-space, and each harmonic component in $k$-space corresponds to a unit cell.

2.	Randomness. All the wavefront should be considered since $n$ is any integer in Eq. (\ref{eq:e3}).

3.	Stable condition Eq. (\ref{eq:11}). In contrary to the general case, the condition of sub-wavelength is not sufficient to mimic a continuous phase distribution, and the stable condition should also be accomplished.

4.	Evanescent waves. Although $\left| k \right| \le {k_0}$  (wavevector in free space) should be satisfied for all propagating waves, solutions to Eq. (\ref{eq:e3}) with large values of $n$ must exist due to the principle of the phase uncertainty.

\section{APPLICATIONS IN METALENS}
\subsection{ARBITRARY MULTI-FOCI METALENS}

\begin{figure}[t]
\centerline{\includegraphics[width=9cm]{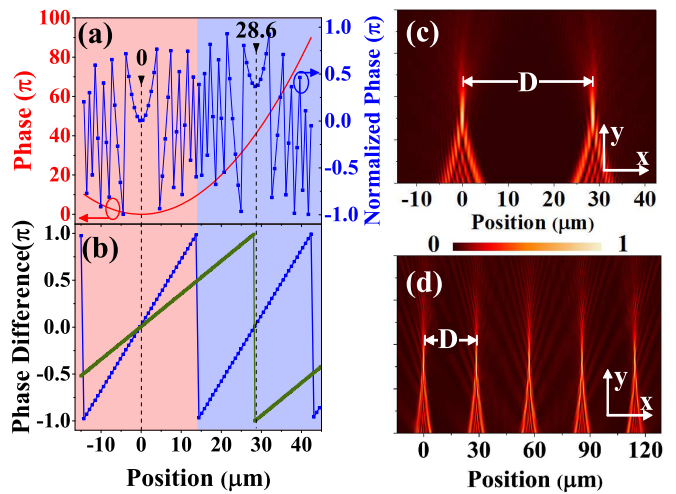}}
\caption{(color online) Multi-foci metalens with a lattice size of 700 nm, focal length of 20 $\mu$m operating at a wavelength of 1 $\mu$m. (a) Parabolic and its normalized phase distribution of discrete unit cells along $x$-axis. The red and blue areas indicate different regions of the metalens that serve as independent foci. (b) Phase difference of the adjacent unit cells with a lattice size of 700 nm (blue dotted line) and 350 nm (green dotted line). (c) Amplitude distribution of a two-foci metalens. (d) Amplitude distribution of a five-foci metalens. }
\label{fig2}%
\end{figure}
The necessity of stable condition can be interpreted through a metalens. It is well known that a hyperbolic phase profile equals to a parabolic phase distribution under the paraxial approximation, which is commonly used to simplify the design of a metalens \cite{39}. However, if the phase profile is defined as $\varphi (x) = {k_0}{x^2}/2f$, a two-dimensional multi-foci metalens can be achieved when $x$ covers a large range, as shown in Fig. \ref{fig2}. The designed metalens has a unit size of 700 nm and focal length of 20 $\mu$m operating at a wavelength of 1 $\mu$m. Interestingly, the original phase profile and the normalized phase profile are both aperiodic [Fig. \ref{fig2}(a)]; however, the focusing profile is periodic. We designed this multi−foci lens to demonstrate the stable condition, which is different with the theory in the Ref. \cite{40}. Evidently this metalens breaks the stable condition $\Delta {\phi _i} \simeq \rm{C}$. As illustrated in Fig. \ref{fig2}(b), the phase difference between two adjacent unit cells varies periodically from $-\pi$ to $\pi$, which cannot be considered close to a constant. To theoretically derive this unique phenomenon, let us first consider the phase difference between two adjacent unit cells. Assuming the size of the unit cells is fixed as $\Lambda$, the phase difference is $\Delta \varphi (m\Lambda ) = \varphi ((m + 1)\Lambda ) - \varphi (m\Lambda ) = {k_0}{\Lambda ^2}(1 + 2m)/2f$, where $m$ is an integer representing the $m$th unit cell. Thus, the localized wave vector is ${k_m} =  - \Delta \varphi /\Lambda  =  - {k_0}\Lambda (1 + 2m)/2f$. If the focus of the system is periodic, the localized wave vector should also be periodic. Considering the randomness of the phase, we can obtain the following formula:
\begin{align}
{k_{m + a}} = {k_m} - \frac{{2n\pi }}{\Lambda },
\label{eq:12}
\end{align}
where $a$ is also an integer representing the number of unit cells for each period. Equation (\ref{eq:12}) can be precisely solved:
\begin{align}
a\Lambda  = \frac{{2n\pi f}}{{{k_0}\Lambda }}.
\label{eq:13}
\end{align}
Thus, when $n=1$ the minimum period of the foci can be calculated as 28.6 $\mu$m from Eq. (\ref{eq:13}), which is in agreement with the simulated results in Fig. \ref{fig2}(c). Another interesting phenomenon is that although the size of the unit cell is less than the wavelength, the diffracting field is much different from that of a continuous phase distribution. According to Eq. (\ref{eq:13}), even when $\Lambda  = \lambda /10$, the period of the foci is 10$f$, which is still not converged. Furthermore, Eq. (\ref{eq:13}) can be reformed as:
\begin{align}
\Lambda (\frac{{a\Lambda }}{f}) = n\lambda.
\label{eq:14}
\end{align}
Compared this equation with the grating equation $\Lambda \sin \theta  = n\lambda $, the designed multi-foci metalens is indeed a beam splitter just like a grating. However, for the gratings, $\left| {\sin \theta } \right| \le 1$ is maintained; whereas, for the metalens, ${a\Lambda }/{f}$ can theoretically be arbitrarily designed. The numbers of foci are linearly proportional to the size of the metalens according to Eq. (\ref{eq:13}). A five-foci metalens is shown in Fig. \ref{fig2}(d), with the same foci distance of 28.6 $\mu$m. It should be noticed that the amplitudes of the foci are equal due to the periodicity of the localized wave vector originating from the phase gradient of the metalens. The metalens can serve as a generalized grating with the same strength for all the diffracting orders.

\subsection{CONVEX-CONCAVE DOUBLE LENS}

\begin{figure}[t]
\centerline{\includegraphics[width=\columnwidth]{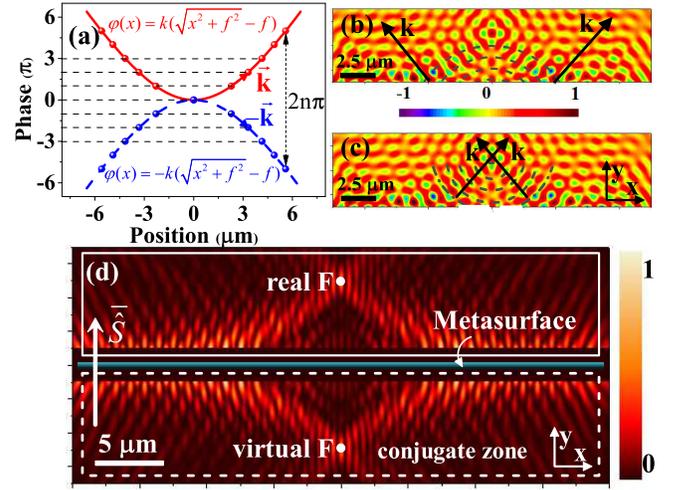}}
\caption{(color online) Designed convex-concave double lens with focal length of 5 $\mu$m operating at a wavelength of 1 $\mu$m. (a) Phase distribution along the $x$-axis (red dots) and its image phase distribution (blue dots). Locations of integral $\pi$ are picked up to locate the unit cells of the lens. (b)$-$(c) Transmitted field profile of the diverged and focused wavefronts. (d) Strength distribution of the transmitted field (white solid box) and effective incident field (white dashed box). The real and virtual focus are both depicted. }
\label{fig3}%
\end{figure}

We also used the proposed theory to design a convex-concave double lens. The unit cells of the metasurface are located at a hyperbolic phase distribution $\varphi (x) = {k_0}(\sqrt {{x^2} + {f^2}}  - f) = n\pi \;(n \in {{\mathbb{N}}})$, as indicated using red dots in Fig. \ref{fig3}(a). It is obvious that these units can generate a focused wavefront [Fig. \ref{fig3}(c)]. However, due to the randomness of the phase, the phase profile can also be projected to another function $\varphi (x) =  - {k_0}(\sqrt {{x^2} + {f^2}}  - f) =  - n\pi \;(n \in {{\mathbb{N}}})$, which can simultaneously achieve a concave lens [Fig. \ref{fig3}(b)]. Similarly, if the blue doted phase profile in Fig. \ref{fig3}(a) is first designed, the red doted one will occur as well. The one-to-one mapping conjugate phase configuration indicates that this device can converge or diverge an incident photon with the same probability. As shown in Fig. \ref{fig3}(d), the amplitudes of real and virtual F are both 0.55 according to the simulation, which implies that convex and concave equally serve the functionality of the device. The transmitted field in the white solid box in Fig. 3(d) is totally mirror-symmetric to the conjugate zone, and this is a direct consequence of the one-to-one mapping conjugate phase configuration.

\begin{figure}[b]
\centerline{\includegraphics[width=\columnwidth]{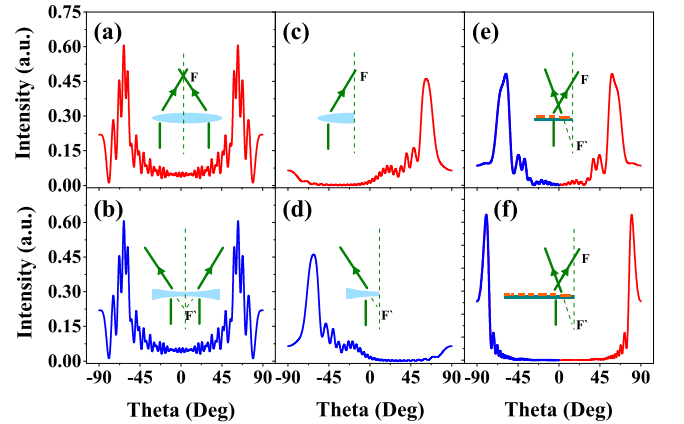}}
\caption{(color online) Calculated $\mathcal{F}$-parameters for (a) focusing metalens, (b) diverging metalens, (c) left side of the focusing metalens, (d) left side of the diverging metalens, (e) left side of the convex-concave double lens with 20 unit cells, (f) left side of the convex-concave double lens with 100 unit cells. Insets: schematics of all the above-mentioned devices. Metalenses in (a)-(d) are designed with a unit size of 500 nm and a focal length of 5 $\mu$m operating at wavelength of 1 $\mu$m.}
\label{fig4}%
\end{figure}

To further examine the exact diffraction patterns of the output field, we calculated the $\mathcal{F}$-parameters in Eq. (\ref{eq:e5}) for the regular focusing metalens, the regular diverging metalens, and the convex-concave double lens, respectively. In Fig. \ref{fig4}, the horizontal ordinate of the graph is calculated through $\theta  = \arcsin (k/{k_0})$, characterizing the divergent angle of output light. The sharp peaks appearing in the lines are caused by the interference of the scattered field, which cannot be eliminated even after improving the calculating accuracy. Interestingly, one cannot recognize a convex and a concave lens from a far distance because a parallel incident light both diverges after passing the focus of the lens. As shown in Figs. \ref{fig4}(a) and \ref{fig4}(b), the $\mathcal{F}$-parameters are the same for a convex and a concave metalens. To distinguish the two lenses, we only analyzed the left sides of the devices [Figs. \ref{fig4}(c) and \ref{fig4}(d)]. For a left-sided convex metalens (30 unit cells utilized), the transmitted light travels toward the right side ($\theta  \ge 0$); whereas for a left-sided concave metalens, the transmitted light travels toward the left side ($\theta  \le 0$). In contrast, $\mathcal{F}$-parameters for the left-sided convex-concave double lens are calculated in Fig. \ref{fig4}(e), which combine Fig. \ref{fig4}(c) with Fig. \ref{fig4}(d), demonstrating that it can simultaneously work as a convex lens and a concave lens. The minor differences between Figs. \ref{fig4}(c)--\ref{fig4}(d) and Fig. \ref{fig4}(e) can be attributed to other orders of diffraction, which primarily exist when $\left| x \right|$ and $n$ is small. For example, with ${x_1} = 0\;(n = 0)$ and ${x_2} =  - 2.29 \;\mu m\;(n = 1)$, $ - \nabla {\varphi _0} = 0.22{k_0}$ can be obtained. Meanwhile, other phase gradients $ - \nabla {\varphi _n} =  - (\Delta \varphi  + 2m\pi )/\Delta x = \{  - 0.22{k_0},0.66{k_0}, - 0.66{k_0}\} $ are permitted due to the randomness of phase. On the contrary, when $\left| x \right|$ or $n$ is large enough, $\Delta x$ can be less than a wavelength and $\left| {2m\pi /\Delta x} \right| > {k_0}$ is maintained for most of the integer $m$. In this situation, only the diffracting order of convex and concave lens can be satisfied and efficiency of each component can approach $\sim 0.5$. The numbers of unit cells are not important for evaluating the functionality of the devices though they can decide the peak position of the $\mathcal{F}$-parameters. The divergent angle increases as the distance from the center of a lens increases. Thus, the $\mathcal{F}$-parameters of an infinite lens should possess two large peaks around ${90^ \circ }$ and ${-90^ \circ }$. In Fig. \ref{fig4}(f), the $\mathcal{F}$-parameters of the convex-concave double lens with 100 unit cells are calculated, which is totally in agreement with the intuitional expects. For all the calculation in Fig. \ref{fig4}, we assume ${T^{MS}}$ to be 300 nm, around $\lambda /3$. Results for different ${T^{MS}}$ values are also displayed in Fig. \ref{fig9} in Appendix E.

\subsection{LARGE NA METALENS}

\begin{figure}[t]
\centerline{\includegraphics[width=\columnwidth]{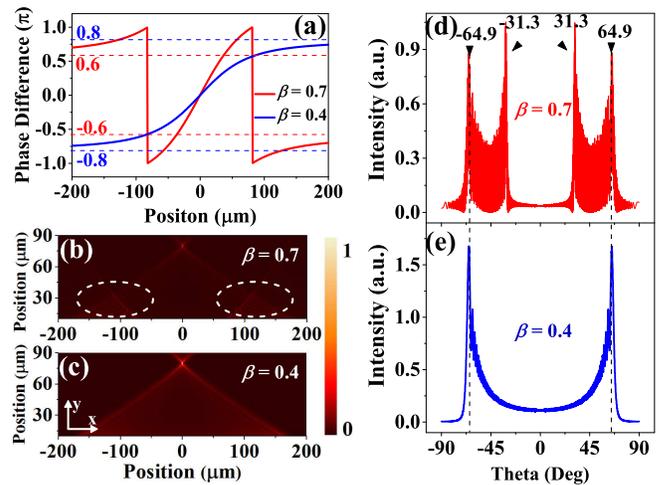}}
\caption{(color online) A hyperbolic metalens with a large NA computed as 0.97 (total size of the lens is 700 $\mu$m). The focal length is 80 $\mu$m, operating at a wavelength of 1 $\mu$m. Size of the unit cells is set to be $\beta \lambda$. (a) Normalized phase difference between two adjacent unit cells with $\beta=0.7$ (red solid line) and $\beta=0.4$ (blue solid line). The dashed lines are asymptotic ones for the corresponding phase differences, on which asymptotic values are marked. (b) Simulated focusing field with $\beta=0.7$, and the white dashed circles indicate zones of high-order diffraction. (c) Simulated focusing field with $\beta=0.4$ without high-order diffractions. Computed $\mathcal{F}$-parameters for (d) $\beta=0.7$ and (e)$\beta=0.4$. }
\label{fig5}%
\end{figure}

For the sub-wavelength hyperbolic metalens, the momentum analysis is also necessary especially when NA is large. Considering a phase profile $\varphi (m\Lambda ) = {k_0}(\sqrt {{{(m\Lambda )}^2} + {f^2}}  - f)$, where $\Lambda$ is size of the unit cell. We can calculate the phase difference between two adjacent unit cells by $\Delta \varphi (m\Lambda ) = {k_0}(\sqrt {{{(m\Lambda  + \Lambda )}^2} + {f^2}}  - \sqrt {{{(m\Lambda )}^2} + {f^2}} )$. When $m$ is large, $\Delta \varphi (m\Lambda )$ has a limit of ${k_0}\Lambda $, independent of the focal length $f$ (derived utilizing the Taylor series ${(1 + x)^a} \approx 1 + ax$), which implies that the hyperbolic phase distribution satisfies the stable condition. Define $\beta  = \Lambda /\lambda $, then $\Delta \varphi (m\Lambda )$ approaches $2\beta\pi$ if $m$ is large enough. As an example, the phase differences for $\beta=0.7$ and $\beta=0.4$ are calculated, as shown in Fig. \ref{fig5}(a). The dashed lines, indicating asymptotic ones for the corresponding phase difference, are exactly same as those of the calculated results with $2\beta\pi$. The localized wave vector can be permitted as propagating waves when it satisfies the condition:
\begin{align}
\left| {\frac{{{k_0}\Lambda }}{\Lambda } + \frac{{2n\pi }}{\Lambda }} \right| \le {k_0},
\label{eq:15}
\end{align}
which leads to:
\begin{align}
\left| {1 + \frac{n}{\beta }} \right| \le 1.
\label{eq:16}
\end{align}
Considering the sub-wavelength condition $0 < \beta  < 1$ is often required, the solution of Eq. (\ref{eq:16}) is $\{ n = 0,\;0 \le \beta  < 1\} $, or $\{ n = -1,\;0.5 \le \beta  < 1\} $. When the NA of the metalens is large enough and $0.5 \le \beta  < 1$, the diffracting order of $ - 1$ will occur. As indicated by the white dashed circles in Fig. \ref{fig5}(b), an obvious diffraction occurs when $\left| x \right| \ge 70$ $\mu$m in the case of $\beta=0.7$. In contrast, in the case of $\beta=0.4$, the high orders of diffraction are suppressed to near field evanescent components and only a focusing wavefront is permitted. An accurate method to evaluate diffraction is to calculate the $\mathcal{F}$-parameters, as shown in Figs. \ref{fig5}(d) and \ref{fig5}(e). We can see that the $\mathcal{F}$-parameter for $\beta=0.7$ has more peaks (diffracting wave vectors) around $ \pm {31.3^ \circ }$. However, the calculated $\mathcal{F}$-parameter for $\beta=0.4$ is more smooth and only has two main peaks at around $ \pm {64.9^ \circ }$, which just overlap with the corresponding peaks for $\beta=0.7$.

\section{CONCLUSION}

In conclusion, we have deduced an enhanced diffraction theory to evaluate the far field for an arbitrary UCWC metasurface, which is based on estimating the width of the source in a unit cell and performing Fourier transform of the unit cell's responding function. We proposed four guidelines to design metasurfaces, especially when the sub-wavelength condition is no longer sufficient to fit a continuous phase distribution. According to the guidelines, the theory has been employed in applications such as: (I) an arbitrary multi-foci lens; (II) a convex-concave double lens; (III) a lens with a large NA preventing undesired diffraction orders. From the theoretical prediction as well as the computational results, the diffracting approach extended to arbitrary phase distributions has been demonstrated to be a powerful tool for guiding the design of multifunctional metasurfaces. Our approach provides a wide platform for designing tailored multifunctional, tunable, especially large-area metasurfaces in practical applications.

\begin{center}
\textbf{ACKNOWLEDGMENTS}
\end{center}
This work was supported by the National Key Research and Development Program of China (2016YFA0301102), the Natural Science Foundation of China (11574163 and 61378006), the Program for New Century Excellent Talents in University (NCET-13-0294), and the 111 project (B07013).

\begin{center}
\textbf{APPENDIX A: CONSERVATION OF MOMENTUM FOR BASIC OPTICAL ELEMENTS}
\end{center}

In quantum optics, momentum of a photon is linearly related to wave vector: ${\bf{p}} = \hbar {\bf{k}}$. Thus, conservation of momentum also means conservation of wave vectors. The law of reflection states that ${\theta _r} = {\theta _i}$, which can also be written as
\begin{align}
{k_0}\sin {\theta _r} = {k_0}\sin {\theta _i} \Leftrightarrow {k_{\parallel r}} = {k_{\parallel i}}.\tag{A1}
\label{A1}
\end{align}

The law of refraction states that ${n_t}\sin {\theta _t} = {n_i}\sin {\theta _i}$, which also implies
\begin{align}
{k_{\parallel t}} = {k_{\parallel i}}.\tag{A2}
\label{A2}
\end{align}

The generalized Snell's law states that ${n_t}\sin {\theta _t} - {n_i}\sin {\theta _i} = {{(\lambda d\Phi )} \mathord{\left/
 {\vphantom {{(\lambda d\Phi )} {(2\pi dx)}}} \right.
 \kern-\nulldelimiterspace} {(2\pi dx)}}$ \cite{6}, and it can be rewritten as
\begin{align}
{k_{\parallel t}} - {k_{\parallel i}} = {k_\Lambda },\tag{A3}
\label{A3}
\end{align}
where ${k_\Lambda } = d\Phi /dx$ is the phase gradient along the surface of the metasurface.

The equation of gratings is $d(\sin {\theta _i} + \sin {\theta _m}) = m\lambda $, which can also be written as
\begin{align}
{k_{\parallel i}} + {k_{\parallel t}} = m{k_d},\tag{A4}
\label{A4}
\end{align}
where ${k_d} = 2\pi /d$ is the reciprocal lattice of the grating.

From Eqs. (\ref{A1})$-$(\ref{A4}), we can see that the law of reflection, regular or generalized Snell's law, and equation of gratings, all result from a single principle--conservation of wave vectors along the surface of the device--which motivated us to use $\mathcal{F}$-parameters to characterize a flat optical elements--metasurface.

\begin{center}
\textbf{APPENDIX B: THEORY TEST}
\end{center}

\begin{figure}[!htb]
\centerline{\includegraphics[width=9.5cm]{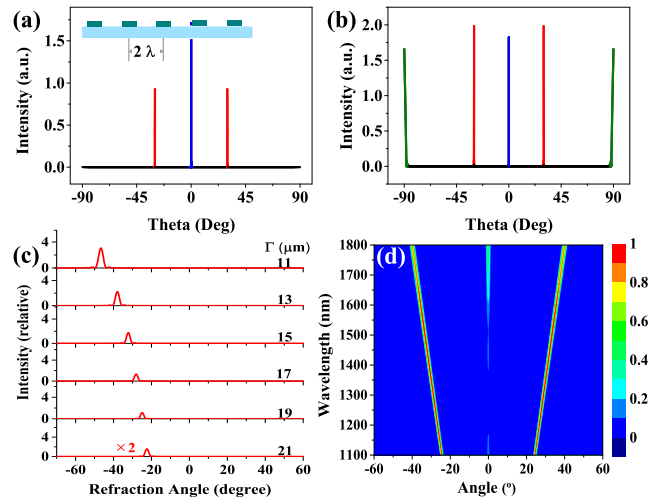}}
\caption{(color online) Computed $\mathcal{F}$-parameters for the metasurface grating with an incident wavelength of 1 $\mu$m and unit cell of 2 $\mu$m. (a) $T^{MS}$ = 900 nm. Inset: schematic of the grating's configuration. (b) $T^{MS}$ = 300 nm. (c) Calculated first order anomalous refraction based on our theory to compare with Fig. 3(c) in Ref. \cite{6}. In the calculation, $T^{MS}$ is set as 1.3 $\mu$m (average size of all the V−shaped antennas taken from Ref. \cite{6}). (d) Calculated scattered intensity based on our model to compare with Fig. 1(c) in Ref. \cite{37}. In the calculation, $T^{MS}$ is set as 1.12 $\mu$m (total size of the three waveguides).}
\label{fig6}
\end{figure}

We utilized a simple metasurface grating to test our theory. In Fig. \ref{fig6}(a)$-$\ref{fig6}(b), the metasurface possesses unity transmittance and same phase delay in each unit cell. The size of a unit cell is set as $\Lambda  = 2\lambda $ ($\lambda$ is the incident wavelength). It is evident that the transmitted light can be scattered to higher orders of diffraction, $\{ 0,\pm\frac{ 2\pi }{\Lambda },\pm\frac{4\pi}{\Lambda },...\} $, corresponding to diffracting angles of $\{ 0, \pm {30^ \circ }, \pm {90^ \circ }\} $ (higher orders become evanescent waves).

Figure \ref{fig6}(a)$-$\ref{fig6}(b) depicts exactly the diffracting orders the metasurface can provide under the responding function (transmission or reflectance) ${t_i}(x) = {\rm{rect[}}(x - {x_i})/{T^{MS}}]{e^{ - i\phi }}$. The blue lines correspond to the zeroth order diffraction, while the red lines indicate first order diffractions. It can be seen that the width of the source in a unit cell, $T^{MS}$, mainly affects the relative strength of each diffracting order. In addition, with $T^{MS} = 300$ nm, the second order diffraction at nearly $ \pm {90^ \circ }$ occurs although diffractions beyond $ {90^ \circ }$ evanesce and cannot be detected from far field. For a traditional grating, the responding function is exactly a rectangular function, while for an arbitrary metasurface, the factor ${\rm{rect}}(x/{T^{MS}})$ is an equivalent estimation of a unit cell's locality.

To test our theoretical model, we compared the experimental results in Ref. \cite{6} with our calculated model. In our calculation, $T^{MS}$ is set as 1.3 $\mu$m by averaging four basic V−shape antennas in Ref. \cite{6}. The total size of the metasurface in the calculation is 230 $\mu$m. The phase and intensity of each antenna are all taken from the reference. Specifically, besides the refraction angle, the height and width of the peaks in Fig. \ref{fig6}(c) coincide with those in Fig. 3(c) from Ref. \cite{6}. We also compared the simulated results in Ref. \cite{37} with our calculated model, as shown in Fig. \ref{fig6}(d). The main peaks of the scattered intensity are in agreement with the results in Ref. \cite{37}, and the relative intensity also make sense. The width of the diffraction peaks in periodic structures is determined by the total number of the unit cells, which is not mentioned in Ref. \cite{37}. We therefore did not compare the width of these peaks with the results in Ref. \cite{37}.

\begin{figure}[!htb]
\centerline{\includegraphics[width=\columnwidth]{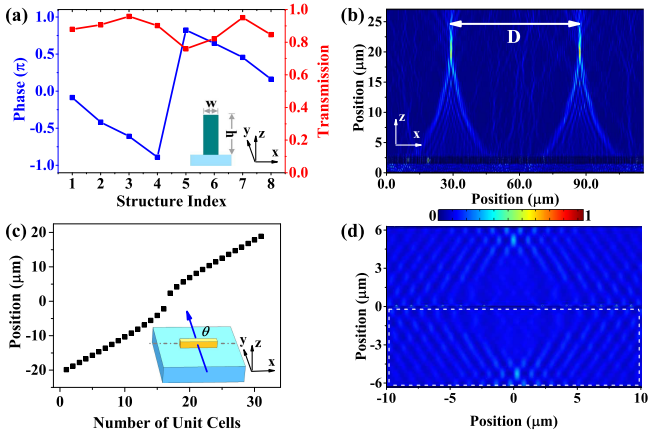}}
\caption{(color online) (a) Phase and transmission of the eight silicon sub−wavelength structure used in the simulation. Inset: Schematic dimension of each silicon structure. (b) Simulated two−foci metalens with the structures in (a). (c) Positions of each unit cell used in the simulation to obtain the convex−concave double lens. Inset: Gold antenna with length of 230 nm, width of 40 nm, and thickness of 40 nm. Orientation angle $\theta$ is fixed at $90^{\circ}$ to get a Pancharatnam–Berry phase of 2$n\pi$. (d) Intensity distribution of the convex−concave double lens.}
\label{fig7}
\end{figure}

The designed metasurfaces in the main text can be easily achieved. As an example, we used a two dimensional simulation with a dielectric metasurface to acquire a similar result with that of Fig. \ref{fig2}(c), as shown in Figs. \ref{fig7}(a)$-$\ref{fig7}(b). The lattice size of the metasurface is 450 nm with a focal length of 20 $\mu$m, operating at a wavelength of 1310 nm and the incident light is polarized along $\emph{y}$ direction. The height of the silicon structure is 1033 nm, and width w of the silicon is 95 nm, 120 nm, 135 nm, 160 nm, 205 nm, 250 nm, 300 nm, and 390 nm, respectively. The refraction index of silicon is set as 3.45. The simulated distance of two foci is D = 58.22 $\mu$m, and the calculated distance according to Eq. \ref{eq:13} is 58.22 $\mu$m as well. We also used a plasmonic metasurface to achieve the convex−concave double lens. The positions of each gold antenna are depicted in Fig. \ref{fig7}(c), and the intensity distribution in Fig. \ref{fig7}(d) is consistent with that in Fig. \ref{fig3}(d).

\begin{center}
\textbf{APPENDIX C: $\mathcal{F}$-PARAMETERS FOR A TWO-DIMENSIONAL METASURFACE}
\end{center}

The responding function of the $i$th unit cell located at $(x_i, y_i)$ can be written as
\begin{align}
{T_i}(x,y) = \left| {{t_i}} \right|{\rm{rect}}(\frac{{x - {x_i}}}{{{T_{xi}^{MS}}}}){\rm{rect}}(\frac{{y - {y_i}}}{{{T_{yi}^{MS}}}}){e^{ - i{\phi _i}}}.\tag{C1}
\label{C1}
\end{align}

By performing a two-dimensional Fourier transform, the diffracting field in $k$-space is
\begin{align}
\mathcal{F}_i({k_x},{k_y}) = &\frac{{\left| {{t_i}} \right|}}{\pi }{e^{ - i({\phi _i} + {k_x}{x_i} + {k_y}{y_i})}} \times \nonumber \\
&\frac{{\sin ({k_x}{T_{xi}^{MS}}/2)}}{{{k_x}}}\frac{{\sin ({k_y}{T_{yi}^{MS}}/2)}}{{{k_y}}}.
\tag{C2}
\label{C2}
\end{align}

Thus, the total responding function and diffraction pattern should be
\begin{align}
T({\bf{r}}) = \sum\limits_i {\left| {{t_i}} \right|\prod\limits_{v = x,y}^{} {{\rm{rect}}(\frac{{v - {v_i}}}{{{T_{vi}^{MS}}}})} {e^{ - i{\phi _i}}}},\tag{C3}
\label{C3}
\end{align}
\begin{align}
\mathcal{F}({\bf{k}}) = \sum\limits_i {\frac{{\left| {{t_i}} \right|}}{\pi }{e^{ - i({\phi _i} + {\bf{k}} \cdot {{\bf{r}}_i})}}\prod\limits_{v = x,y}^{} {\frac{{\sin ({k_v}{T_{vi}^{MS}}/2)}}{{{k_v}}}} }.\tag{C4}
\label{C4}
\end{align}

\begin{center}
\textbf{APPENDIX D: CALCULATING METHODS}
\end{center}

Field profiles in Figs. \ref{fig2}(c)-–\ref{fig2}(d), \ref{fig3}(b)–-\ref{fig3}(d), \ref{fig5}(b)-–\ref{fig5}(c) in the main text are calculated via commercial software MATLAB with each unit cell of the metasurface regarded as a secondary source that can radiate a spherical wave. Based on theoretical and experimental research from other groups \cite{6,7,8,9,10,11,12,13,14,15,16,17,18,28,29,30,31,32,33,36}, the main properties of a metasurface is decided by the strength of the output light and the phase of each unit cell, despite the unit cell's radiating type. Thus, we use a spherical wave as a secondary source to test the validity of the field profile.

Specifically, the conjugate zone in Fig. \ref{fig3}(d) is computed through deduction as follows:

\begin{figure}[!htb]
\centerline{\includegraphics[width=5cm]{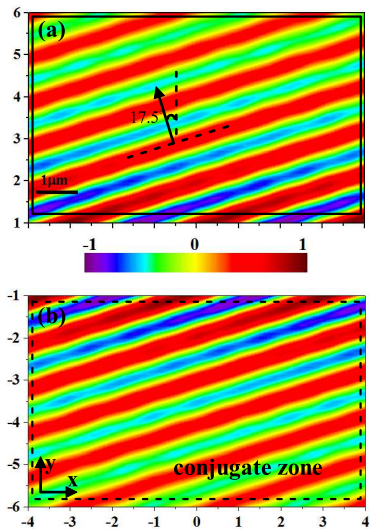}}
\caption{(color online) (a) A propagating plane wave and (b) rebuilt incident wave depicted in a dashed box. }
\label{fig8}%
\end{figure}
Wavefront is the contour map of the wave's phase, and wave vector is the gradient vector of this map, written as ${\bf{k}} =  - \nabla \varphi $. Thus, the wave will propagate along ${\bf{k}}$. If one want to recovery the incident wave, we can just reverse the direction of ${\bf{k}}$ such that $ - {\bf{k}} = \nabla \varphi  =  - \nabla ( - \varphi )$. As shown in Fig. \ref{fig8}(a), a plane wave is generated by a metasurface with a phase distribution of $ - 0.3{k_0}r$ located along $y = 0$. The wave propagates at $-17.5^\circ$, as computed with $\arcsin ( - 0.3)=- {17.5^ \circ }$. Figure \ref{fig8}(b) is computed by reversing the phase distribution as $0.3{k_0}r$ to rebuild the incident wave.

\begin{center}
\textbf{APPENDIX E: DIFFRACTING PATTERNS WITH DIFFERENT VALUES OF $T^{MS}$}
\end{center}

\begin{figure}[!htb]
\centerline{\includegraphics[width=\columnwidth]{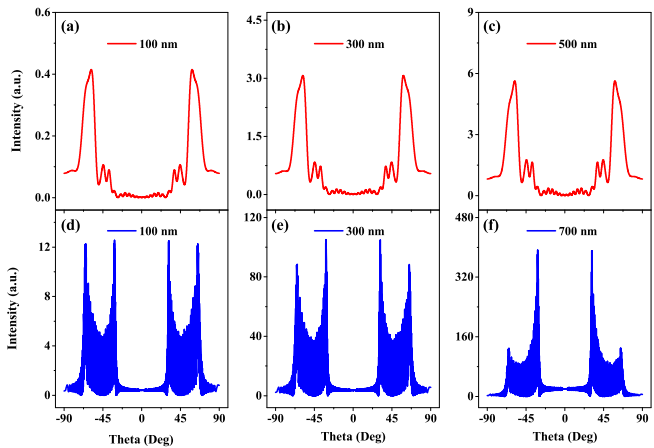}}
\caption{(color online) Calculated $\mathcal{F}$-parameters for the convex-concave double lens designed in Figs. \ref{fig2}--\ref{fig3} with (a) $T^{MS}$ = 100 nm, (b) $T^{MS}$ = 300 nm, (c) $T^{MS}$ = 500 nm. Calculated $\mathcal{F}$-parameters for the metalenses designed in Fig. \ref{fig5} with (d) $T^{MS}$ = 100 nm, (e) $T^{MS}$ = 300 nm, and (f) $T^{MS}$ = 700 nm. }
\label{fig9}%
\end{figure}

$\mathcal{F}(k) = \sum\limits_i {\frac{{\left| {{t_i}} \right|}}{\pi }\frac{{\sin (k{T_i^{MS}}/2)}}{k}} {e^{ - i({\phi _i} + k{x_i})}}$ degenerates to $\mathcal{F}(k) \approx \sum\limits_i {\frac{{{T_i^{MS}}\left| {{t_i}} \right|}}{{2\pi }}} {e^{ - i({\phi _i} + k{x_i})}}$ when the width parameter $T^{MS}$ is sufficiently small and the strength of  $\mathcal{F}$-parameters is approximately proportional to the size of $T^{MS}$. This also means that when $T^{MS}$ is sufficiently small, the diffracting patterns remain the same regardless of the value of $T^{MS}$. However, when $T^{MS}$ is large (still smaller than a wavelength), it will affect $\mathcal{F}_i$ as a sine function; the corresponding results can be seen in Fig. \ref{fig6}(a)--\ref{fig6}(b). We also plotted different diffracting patterns of the designed convex-concave double lens for different $T^{MS}$, as shown in Figs. \ref{fig9}(a)--\ref{fig9}(c). The incident wavelength is 1 $\mu$m and the size of the unit cell is 500 nm. All the values of $T^{MS}$ are chosen as ${T^{MS}} < \Lambda  = 500$ nm, and all the diffracting patterns are almost the same in this case. As for metalenses designed in Fig. \ref{fig5},  $\mathcal{F}$-parameters differ when varying sizes of $T^{MS}$ from 100 nm to 700 nm, as shown in Figs. \ref{fig9}(d)--\ref{fig9}(f). The position of each diffractive peak remains the same, while the amplitude varies for different values of $T^{MS}$, and this result fits with the diffracting patterns of a metasurface grating, as stated in Appendix B.

\end{document}